\input{aipcheck}

\documentclass[
    ,final            
  ]
  {aipproc}

\layoutstyle{6x9}

\begin{document}

\title{Quark-gluon densities in the nuclear fragmentation region in heavy ion collisions at LHC
}

\classification{25.75.Nq}
\keywords      {fragmentation, heavy ions}

\author{Mark Strikman}{
  address={104 Davey lab, Penn State University, University Park, PA 16802, U.S.A.}
}

\begin{abstract}
At the LHC, the leading partons in the nuclei are expected to 
interact with the maximal possible strength - black disk limit  - up to transverse momenta of the order of few GeV. We demonstrate that in this limit the densities of  the quark - gluon systems produced in the central AA collisions in the nucleus fragmentation regions should exceed $300 GeV/fm^3 $ which is 
at least as high as the densities discussed for the central region. Experimental signatures of such a regime are also discussed.

\end{abstract}

\maketitle


The focus of most  studies of the quark-gluon state produced in  heavy ion collisions is the central region where one expects generation of high gluon densities at sufficiently high energies. The first  estimates of the hadron matter densities produced in the nuclear fragmentation region 
were presented  in \cite{Larry} in the framework of the soft  hadronic dynamics. The first estimate of the quark-gluon densities in this kinematics was presented within the framework of 
the onset of the black disk limit (BDL) of QCD in \cite{prl} which found   densities  at least a factor of ten larger. 
The purpose of this talk is review and update analysis the analysis of \cite{prl}.

The starting point of \cite{prl} is the observation \cite{martin}  that the transverse momenta of partons propagating through a high gluon density medium should become much larger than the scale of soft interactions. This is due to  the possibility for a quark with given large $x_1$ to interact with partons with  very small $x_2=4p_t^2/x_1s$ where $p_t$ is the resolution scale.
 As a result, for the case of a gluon  with $x\ge 10^{-2}$ propagating through the center of the heavy nucleus  we find an average of $p_t \ge 4 GeV/c$ at the LHC energies, see review in \cite{annual}.

 Let us consider nucleus - nucleus  scattering  in the rest frame of one of the nuclei.
First let us determine the emission angle,  $\theta$,  of the parton belonging to the nucleus which was at rest. Since the energy losses are small, the light-cone fraction carried by the parton is approximately conserved:
\begin{equation}
(E_i-p_i^z )=xm_N,
\end{equation}
leading to 
$p_z=(\mu^2+p_t^2)/2xm_N -xm_N/2 \approx p_t^2/2xm_N$.
Here in the last step  we have neglected $\mu^2$  compared to  
$p_t^2$ which is legitimate in the leading order. Since $\mu^2\geq 0$,
neglected terms would increase $p_z$ making the emission angles, $\theta$,
even smaller. Thus, in the BDL the angles $\theta\simeq p_t/p_z\sim 2x m_N /p_t$ are small. So,  the length  of the  produced  
wave package is reduced from the naive value of $2R_A$ by the large factor $S=1/(1-\cos\theta)\approx p_t^2 / 2x^2 m_N^2$.

However, we must also take into account that the products of the nucleon fragment as a whole move forward in the target rest frame. Since the knocked out partons carry practically the whole light cone fraction of the nucleon,  the mass squared of the produced system, $M^2$ and its longitudinal momentum, $p_z$ can be determined from $
(\sqrt{M^2+p_z^2}-p_z)/m_N=1, M^2=\sum_i {p_{i,~t}^2/ x_i}, $ where $x_i,p_{i,~t}$ are light cone variables for produced partons.
Hence, $p_z=M^2/2m_N$, and the Lorentz factor $\gamma=E/M=\sqrt{M^2+(M^2/2m_N)^2}/M \approx M/2m_N$.
As a result we find the total reduction in the volume:
\begin{equation}  
D=(2m_N/M)\cdot  <p_t^2/ 2m_N^2 x^2>.
\label{9}
\end{equation}
Since the energy of the system in its rest frame is $M$, we find for the overall enhancement as compared to the nuclear density:
\begin{equation}  
R_E= {1\over N_q+N_g} \sum_i {p_{i\,t}^2\over  m_N^2 x_i^2}.
\end{equation}
To illustrate the  dependence of $R_A$ on the total number of involved partons, N,
 and on average transverse momenta we can take all $x_i$ and all $p_{i\,t}$. In this case  $D=Np_t/m_N, \,  R_E=D^2$,
  and energy density depends quadratically  on the average transverse momentum of partons.

Our estimates indicate that at LHC for the gluons with $x\ge 0.0.5$, $\left<p_{g\,t}^2\right> \ge 16 GeV^2$ (and growing with increase of x), and that for quarks $\left<p_{q\,t}^2\right>$ is a factor of two smaller \cite{annual}.
Taking for  illustration $N_q=3, N_g=6, x_q=1/6, x_g=1/12$ and $\left<p_{g\, t}^2\right> = 20 GeV^2$ we find  $R_E= 2300$. This corresponds to 
$$"energy \, \,  density" \sim 370 \, GeV/fm^3,$$
which is at least as large as the one expected for the central region. It 
  is much larger than our initial estimate where a very conservative value of $\left<p_{g\,t}^2\right>$ was taken. 
  
If we assume the proximity of BDL  at  RHIC for
the  fragmentation region for   $p_t\sim 1 GeV/c$,
we find quark-gluon energy densities $\sim 10 \,GeV/fm^3$.
These densities are at least a factor of 10
higher than in \cite{Larry} due to much larger release of energy in
the BDL and due to significantly  larger  longitudinal compression
of the interaction volume.
Our estimate neglects the  conversion of the released partons into hadrons 
before they reach the back edge of the fragmenting nucleus. 
If $p_t$ generated in the collision is small enough, this effect may become important.

Using the logic similar to the one we used for estimating hadron formation in the color transparency phenomenon \cite{Farrar} we can estimate the distance over which a parton (not interacting with a medium) converts to hadrons. One finds 
\begin{equation}
l_{coh} = 2p_q/\Delta M^2 \,
\end{equation}
where $\Delta M^2$ is the mass gap between two lowest hadrons with the same quantum numbers.
Numerically,   $l_{coh} \approx 0.3\div 0.4\, fm
\cdot p_q [GeV]$, corresponding for $p_q=4 GeV$.
This is substantially larger than the expected interaction length (see below) and hence the evolution of the imploded system should be determined mostly by partonic  rather than hadronic interactions.

The difference in
average x's of quarks and gluons leads to a different direction of the
flow of the quarks and gluons in the center of mass frame of the produced
system. For the above numerical example,  $k_z/k_t \sim 0.7 $ for    
quarks and  $\sim -0.25$ for gluons. Obviously, this pattern       
will enhance the interactions of quarks and gluons at the next stage   
of the interactions, making equilibration more likely.

It follows from the above analysis
that at LHC in the first stage of
collisions a strongly compressed hot  quark-gluon state of the 
ellipsoidal shape is formed with the small principal
axis of $\sim 0.5 \, fm$ and density  $\rho \geq 50$ partons per $fm^3$.
At the higher rapidity end, this ellipsoid boarders essentially parton free
space; on the end close to central rapidities, it boarders a hot 
$q\bar q g$ state.
The scattering length for parton $i$ can be estimated as $l_i=1/(\sum_j
N_j \sigma_{ij})$, corresponding to the scattering length being smaller
than 0.5 fm for $\sigma\ge 0.5 mb$. To estimate the interaction cross
section, 
we note that the average invariant energy $s\approx 2p_t^2\sim 32
GeV^2$. The  initial stage of reinteractions  certainly is a highly
non-equilibrium process. Nevertheless to do a perturbative estimate 
 we can conservatively introduce a cutoff
on the momentum transfer $p\sim {\pi\over 2}\rho^{-1/3}$, leading to the
leading order estimate for the gluon - gluon cross section $\geq 1
mb$. Nonperturbative effects, which remain strong in the gluon sector
up to $ s \sim 10$ GeV$^2$, are likely to increase these interactions further. 
Consequently, we expect partons to rescatter strongly at the second
stage, though much more detailed modeling is required 
to find out whether the system may reach thermal equilibrium.

The large angle rescatterings of partons
will lead to production of partons at higher rapidities
and re-population of the cool region. In particular, 
two gluons have the right energies to produce,  via gluon fusion
 $c\bar c, b\bar b$
pairs and in particular $\chi_c,  \chi_b$-mesons with rather
large transverse momenta and $x_F \sim 2x_g \sim 0.1$. 
Also,  leading photons can
be produced in the $q g \to \gamma q$ subprocesses, though in
difference from the central region the $q\bar q \to \mu^+\mu^-$
production will be
suppressed due to the lack of antiquarks.  Another high density effect
is the production of leading 
nucleons via recombination of quarks 
with subsequent escape to the cool region. Hence we expect a rather
paradoxical situation that the production of leading hadrons in $AA$
collisions will be stronger than in the central pA collisions.

To summarize, we have demonstrated that the onset of the BDL
 in the interactions in the target fragmentation region which is
likely at LHC for 
a large range of virtualities,  will lead to the formation of   a
new superdense initial state in the nuclear fragmentation region with densities
exceeding nuclear densities at least by a factor of 2000.
Our reasoning is however insufficient to demonstrate whether
thermalization processes will be strong enough for the system to reach
equilibrium necessary for  formation of
metastable states.


\begin{thebibliography}{9}
\bibitem{Larry}R.~Anishetty, P.~Koehler and L.~D.~McLerran,
Phys.\ Rev.\ D {\bf 22}, 2793 (1980).
\bibitem{prl}L.~Frankfurt and M.~Strikman,
  Phys.\ Rev.\ Lett.\  {\bf 91}, 022301 (2003)  [arXiv:nucl-th/0212094].
\bibitem{martin}L.~Frankfurt, V.~Guzey, M.~McDermott and M.~Strikman,
Phys.\ Rev.\ Lett.\  {\bf 87} (2001) 192301
[arXiv:hep-ph/0104154].
\bibitem{annual}L.~Frankfurt, M.~Strikman and C.~Weiss,
Ann.\ Rev.\ Nucl.\ Part.\ Sci.\  {\bf 55}  (2005) 403.
\bibitem{Farrar}
  G.~R.~Farrar, H.~Liu, L.~L.~Frankfurt and M.~I.~Strikman,
  Phys.\ Rev.\ Lett.\  {\bf 61}, 686 (1988).
\end{thebibliography}
\end{document}